\documentclass[prl,aps,preprint]{revtex4}
\usepackage[dvips]{graphicx}

\begin{document}
\title{Origin of attraction between likely charged hydrophobic and hydrophilic walls confining  near-critical
 binary aqueous mixture with  ions}
\author{Faezeh Pousaneh and Alina Ciach}
\affiliation{Institute of Physical Chemistry,
 Polish Academy of Sciences, 01-224 Warszawa, Poland}

\begin{abstract} 
 Effect of ionic solute on a near-critical binary aqueous mixture confined
between charged walls with different adsorption preferences is considered within a simple density functional theory.
 For the near-critical system containing small amount of ions a Landau-type functional is derived based on the assumption 
that the correlation, $\xi$, and the Debye screening length, $\kappa^{-1}$, are both much larger than the molecular size. 
The corresponding approximate Euler-Lagrange equations are
solved analytically for ions insoluble in the organic solvent. Nontrivial concentration profile of the solvent
 is found near the charged 
hydrophobic wall as a result of the competition between the short-range attraction
 of the organic solvent and the electrostatic
 attraction of the hydrated ions. Excess of water may be present near the hydrophobic surface for some range of 
 the surface charge and $\xi\kappa$. As a result, the effective potential  
between the hydrophilic and the hydrophobic surface can be repulsive far from the critical point, then attractive and again
repulsive when the critical temperature is approached, in agreement with the recent experiment
 [Nellen at.al., Soft Matter {\bf 7}, 5360 (2011)]. 
\end{abstract}
\maketitle
Near-critical binary mixture confined in a slit induces effective attraction or repulsion between the confining walls if 
adsorption preferences of the two walls are the same or opposite respectively \cite{krech:99:0,nature,gambassi:10:0}. 
The range of this so called thermodynamic 
Casimir force is of order of the bulk correlation length $\xi$. Parallel walls covered by electric charges of the same 
sign repel 
each other. One could thus expect stronger repulsion between likely charged walls with opposite adsorption preferences 
confining 
near-critical binary mixture. In striking contrast to the above expectation, 
in the recent experiments~\cite{nellen:11:0}  strong attraction was observed between a charged hydrophobic colloid particle and 
a charged hydrophilic substrate for some range of temperatures and concentrations of a hydrophilic salt added to the solution. 
Effective interactions between the colloid particles separated by  distances much smaller than their radii are similar to the
interactions between planar surfaces. Possibility of changing these interactions  from attraction to repulsion by minute 
changes of temperature or salinity opens possibilities for designing and  controlling reversible structural changes, 
in particular aggregation or adsorption. It is thus important to understand the mutual influence of the critical adsorption and 
the distribution of ions that leads to the attraction  between the walls instead of the expected repulsion.
 We address this issue in this communication.

We consider a water - organic liquid mixture containing  hydrophilic ions in a slit with selective, charged walls of the area $A\to \infty$, separated by the distance $L\gg 1$ (Fig.1). We choose 
the average diameter of the molecules, $a\equiv 1$, as the length unit and all the corresponding functions are dimensionless. The grand thermodynamic potential of the system
can be written in  terms of the local dimensionless  densities $\rho_i({\bf r})$, 
where $i=1,2,3,4$ for water, oil, $+$ and $-$ ion respectively, in the form~\cite{evans:90:0}
\begin{eqnarray}
\label{Omega}
\Omega&=&-pAL+A\omega_{ex}
\\
\nonumber
&=&\frac{1}{2}\int_{V}d{\bf r}\int_{V} d{\bf r}' \rho_i({\bf r})V_{ij}({\bf r}-{\bf r}') g_{ij}({\bf r}-{\bf r}')
\rho_j({\bf r}')
%\\\nonumber
 +\int_{V}d{\bf r}\rho_i({\bf r})\Big(V_{i}^s({\bf r})-\mu_i\Big)+U_{el}-TS
\end{eqnarray}
where $p$ is the bulk pressure, the integration is over the system volume $V=AL$, periodic boundary conditions are imposed 
in the directions parallel to the walls, and $S$,  $U_{el}$, $T$ and $\mu_i$  are entropy,   electrostatic energy, 
temperature and chemical potential of 
the $i$-th specie respectively. $V_{ij}$ and $g_{ij}$ are the van der Waals (vdW)
interactions and the pair correlations between the corresponding components respectively, and the summation convention for 
repeated indices is  assumed in the whole communication.
$V_{i}^s({\bf r})$ is the sum of the direct wall-fluid potentials acting on the
 component $i$.  Finally,
\begin{eqnarray}
 \omega_{ex}=  (\gamma_0+\gamma_L)+\Psi(L)
  \end{eqnarray} 
is the excess grand potential per surface area, 
$\gamma_n$  is the surface tension at the $n$-th wall ($n=0,L$), and the effective
 potential  $\Psi(L)$ is  the subject of our study. Because of the translational symmetry in the parallel directions,
 the densities depend only on the distance from the left wall,  $z$.
We make the standard approximation
\begin{eqnarray}
\label{S}
 S/A=-k_b\sum_{i=1}^4\int_0^L dz\rho_i(z)\ln\rho_i(z),
\end{eqnarray}
and the standard assumption \cite{barrat:03:0}
\begin{eqnarray}
\label{Uel}
U_{el}[\phi]/A=\int_0^{L}dz \big [\frac{-\epsilon}{8\pi}(\bigtriangledown\psi)^2
+e\phi  \psi\big]
+e\sigma(0)\psi(0)+
e\sigma(L) \psi(L),
 \end{eqnarray} 
where the electrostatic potential  $\psi$ satisfies the Poisson equation,
\begin{eqnarray}
\label{Poisson}
 \frac{\epsilon}{4\pi}\frac{d^2 \psi(z)}{d z^2}+e\phi(z)= 0,
\end{eqnarray}
$e$ is the elementary charge, $\epsilon$ is the dielectric constant of the solvent, $\sigma(n)$ is the dimensionless 
surface charge density
at the  $n$-th wall, and  
\begin{eqnarray}
 \phi(z)=\rho_3(z)-\rho_4(z) 
\end{eqnarray}
is the dimensionless charge  density.
Compressibility of the liquid can be neglected, and we assume $\sum_{i=1}^4\rho_i=1$.  We choose $\phi$,  
the solvent concentration
$s=\rho_1-\rho_2$,  and the density of ions $\rho_c=\rho_3+\rho_4$ as the three independent variables. 
Bulk equilibrium densities for given $T$ and $\mu_i$
 correspond to the minimum of $-pAL$, and are denoted by $\bar s$ and $\bar \rho_c$. 
In equilibrium $\phi(z) $  and the deviations from the bulk values, 
\begin{eqnarray}
 \vartheta_1(z)=\rho_1(z)-\rho_2(z)-\bar s, \hskip1cm  \vartheta_2(z)=\rho_3(z)+\rho_4(z)-\bar  \rho_c
\end{eqnarray}
 correspond to the minimum of 
$\omega_{ex}[\vartheta_1,\vartheta_2,\phi]=(\Omega[\vartheta_1+\bar s,\vartheta_2+\bar \rho_c,\phi]-\Omega[\bar s,\bar \rho_c,0])/A$ with $\bar s$ and $\bar \rho_c$ fixed.  We choose for $\bar s$ and $\bar \rho_c$ the values corresponding to the critical point.

\begin{figure}
  \includegraphics[scale=0.20]{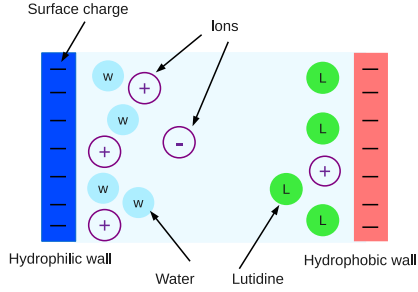}
\caption{Model system consisting of water, organic liquid (for example lutidine) and ions between 
negatively charged hydrophilic (dark, blue) and hydrophobic (light, red) walls.
}
 \end{figure}

Common salts are soluble in water and insoluble in organic solvents.  We thus
 assume that the difference in  the  chemical nature  of the anion and the cation is negligible, 
and postulate the same vdW interactions, $V_{i,3}= V_{i,4}$. From Eq.(\ref {Omega}) it easily follows 
that the vdW contribution  to the internal energy expressed in terms of the new variables is  independent of $\phi$ 
when $V_{i,3}= V_{i,4}$
 \cite{ciach:10:0}. Because  $U_{el}$ is independent of $\vartheta_i$ (see (\ref{Uel})),
in this approximation the vdW and the electrostatic
 contributions to the internal
 energy are decoupled. The coupling is present in the entropic part.  The excess entropy, 
$s_{ex}[\vartheta_1 ,\vartheta_2,\phi]=(S[\vartheta_1+\bar s,\vartheta_2+\bar \rho_c,\phi]-S[\bar s,\bar \rho_c,0])/A$, can be Taylor expanded in terms of
$\vartheta_i(z)$ and $\phi(z)$. For a near-critical mixture with small amount of ions the expansion can be truncated, because   $\vartheta_i(z)$ and $\phi(z)$ are small (except from microscopic distances from the surfaces). 
Using Eq.(\ref{S}) one can verify that the excess entropy contains no terms proportional to $\phi^n\vartheta_1^m$, and the lowest-order mixed term is $\phi^2\vartheta_2$. Thus, the excess grand potential  can be split in two leading-order terms and the correction $\Delta {\cal L}$
\begin{eqnarray}
\label{psi1}
\omega_{ex}[\vartheta_1,\vartheta_2,\phi]\approx {\cal L}_C[\vartheta_1,\vartheta_2] 
+{\cal L}_{DH}[\phi]+\Delta {\cal L}[\vartheta_2,\phi].
\end{eqnarray}
 From the minimum condition for $\omega_{ex}$ it follows that the linear terms vanish, and the dominant terms in
 Eq.(\ref{psi1})
 are quadratic in the fields $\vartheta_i(z)$ and $\phi(z)$.  The second term on the RHS of Eq.(\ref{psi1}) has the form
\begin{eqnarray}
\label{DH}
 {\cal L}_{DH}[\phi]=U_{el}[\phi]/A+\frac{k_B T}{2\bar \rho_c}\int_0^{L}dz\big(\phi^2(z) +O(\phi^4)\big).
 \end{eqnarray} 
 Eqs.  (\ref{DH}), (\ref{Uel}) and (\ref{Poisson}) agree with the Debye-Huckel (DH) theory
for the excess grand potential of
 ions in a homogeneous solvent confined in a slit with charged walls. The first term in Eq.(\ref{psi1})
 is equal to the excess grand potential per unit area for one kind of neutral solute in a two-component solvent, where the excess concentration
of the solvent and the excess solute  density are denoted by
 $\vartheta_1$ and $\vartheta_2$ respectively and the total density is fixed. This is because we assumed
 no difference between the vdW interactions of the anion and the cation - when uncharged, they represent the same species in
 this theory. Close to the critical temperature $T_c$ the fields  $\vartheta_i(z)$
 vary on the length scale  $\xi\propto \mid (T-T_c)/T_c\mid^{-\nu}\gg 1$ with $\nu\approx 0.63$, and the standard coarse-graining procedures leading to the Landau functional can be
 applied~\cite{goldenfeld:03:0}. Our coarse-graining of the first term in Eq.(\ref{Omega}) 
(expressed in terms of the new variables) is based on the Taylor expansion of  $\vartheta_i(z')$  about
 $z'=z$. The excess grand potential is expressed in terms of the fields $\vartheta_i$ and  their derivatives,
 and in terms of the appropriate moments of the vdW interaction potentials,
\begin{eqnarray}
\label{LC}
 {\cal L}_C[\vartheta_1,\vartheta_2]\approx\frac{1}{2}\int_0^{L}dz\Bigg\{\vartheta_i(z)C^0_{ij}\vartheta_j(z)
+\nabla\vartheta_i(z) J_{ij}\nabla\vartheta_j(z)+O(\vartheta_1^4,\vartheta_2^3,\vartheta_1^2\vartheta_2)
\Bigg\} \\
\nonumber +\frac{\vartheta_i(0)J_{ij}\vartheta_j(0)}{2}-h_i(0)\vartheta_i(0)
%\\\nonumber
+\frac{\vartheta_i(L)J_{ij}\vartheta_j(L)}{2}-h_i(L)\vartheta_i(L),
\end{eqnarray}
where 
\begin{eqnarray}
 C^0_{ij}=- J^0_{ij}-T\frac{\partial^2 s_{ex}[ \vartheta_1, \vartheta_2,0]}{\partial \vartheta_i\partial \vartheta_j},
\end{eqnarray}
 $J_{ij}^0=\int d{\bf r} J_{ij}( r)$ and $J_{ij}=\frac{1}{6}\int d{\bf r} J_{ij}( r)r^2$. $- J_{ij}(r)$ represents the vdW
 interactions for $\vartheta_i$ and $\vartheta_j$, and can be obtained from the vdW contribution to Eq.(\ref{Omega}) with
 the densities expressed in terms of the new variables.  
 We shall assume that the interaction ranges $\zeta_{ij}$ defined by $ \zeta^2_{ij} =6J_{ij}/J_{ij}^0$ are all 
 $\zeta_{ij}\approx 1$, and characterize the system by three interaction parameters, $J_{ij}=J_{ij}^0/6$ 
(for the length unit $a\equiv 1$). 
Finally, $h_i(n)$ is the surface field describing  interactions with the $n$-th wall.
 The remaining surface terms result from the compensation for the interactions with the missing fluid
neighbors at the wall. These interactions are included in the bulk term, but should be absent if the wall is present. 
 In Ref.\cite{ciach:10:0} the same functional was obtained  from a lattice model for the four-component mixture.
When the mixture phase separates,  both the solvent concentration and the density of solute  are different in
 the coexisting phases, because the solute is soluble only in water.  Thus, $ {\cal L}_C$ must depend on  both,
 $\vartheta_1$ and $\vartheta_2$. The  critical order parameter is the eigenvector
 of $C^0_{ij}$  corresponding to the eigenvalue vanishing for $T=T_c$. 
 
 In Eqs.(\ref{LC}) and (\ref{DH}) the terms proportional to $k_BT$ represent the leading-order contributions to the excess
 entropy per surface area. The next-to-leading order contribution has the form
\begin{eqnarray}
\label{DL}
\Delta {\cal L}[\vartheta_2,\phi]=-k_B T\int_0^{L}dz \Bigg[\frac{\vartheta_2(z)\phi^2(z)}{2\bar \rho^2_c}+O(\vartheta_2^2\phi^2,\vartheta_2\phi^4)\Bigg]
\end{eqnarray}
and results from the fact that there are more ways of introducing a local difference in the concentrations
 of the anions and the cations, $\phi(z)$, in the regions where there is more ions ($\vartheta_2(z)>0$) than in the 
regions where there is less ions ($\vartheta_2(z)<0$).

The Euler-Lagrange (EL) equations for the functional (\ref{psi1}) -(\ref{DL}),   with the higher order terms in(\ref{LC}), (\ref{DH}) and (\ref{DL}) neglected, take the forms
\begin{eqnarray}
\label{EL1}
 \frac{d^2 \vartheta_i(z)}{d z^2}=M_{ij}\vartheta_j(z) +d_i\phi^2(z) 
\end{eqnarray}
\begin{eqnarray}
\label{EL2}
\frac{d^2 \phi(z)}{d z^2}=\kappa^2\phi(z) +\frac{1}{\bar\rho_c}\frac{d^2 (\phi(z)\vartheta_2(z))}{d z^2}.
\end{eqnarray}
 The charge neutrality condition, $\int_0^Ldz\phi(z)+\sigma_0+\sigma_L=0$, is imposed on $\phi$,  and the boundary conditions   for $\vartheta_i$ are
\begin{eqnarray}
\label{bc0}
\frac{d \vartheta_i(z)}{d z}|_{z=0}-\vartheta_i(0)= H_i(0)
\\
\nonumber
-\frac{d \vartheta_i(z)}{d z}|_{z=L}-\vartheta_i(L)=H_i(L).
\end{eqnarray}
In the above $M_{ij}=(J^{-1})_{ik}C^0_{kj}$, where $(J^{-1})_{ik}$ is the $(i,k)$-th element of the matrix
 inverse to the matrix $J_{ij}$~\cite{ciach:10:0}. 
The remaining parameters are $(d_1,d_2)=-\frac{k_BT}{2\bar\rho_c^2}\Big( (J^{-1})_{1,2},( J^{-1})_{2,2}\Big)$,
 and  $H_i(n)=(J^{-1})_{ij}h_j(n)$. For a hydrophilic (hydrophobic) wall $H_1<0$  ($H_1>0$).

When $\Delta {\cal L}$ in Eq.(\ref{psi1}) is neglected, the Casimir and the electrostatic potentials are
independent contributions to $\omega_{ex}$, and
the EL equations are linear and decoupled (the second terms on the RHS of Eqs.(\ref{EL1}) and (\ref{EL2}) are absent).
 In a semi-infinite system the solutions of the linearized  EL equation
(\ref{EL2}) and (\ref{EL1}) are
\begin{eqnarray}
\label{phi0}
\phi^{(1)} (z)=-\kappa\sigma\exp(-\kappa z)
\end{eqnarray}
where $\kappa^2=\frac{4\pi e^2\bar\rho_c}{k_BT\bar\epsilon} $~\cite{barrat:03:0,goldenfeld:03:0,israelachvili}, and
\begin{eqnarray}
\label{e0}
\vartheta_i^{(1)}(z)=A_i\exp(-z/\xi)+C_i\exp(-\lambda_2 z),
\end{eqnarray}
where $A_i$ and $C_i$ depend linearly on $H_i$. The superscript $(1)$ refers to the solutions of the linearized EL equations.
 In the critical region $\xi\to\infty$ and $\lambda_2\gg1/\xi$, 
therefore the second term on the RHS of Eq.(\ref{e0}) can be neglected. In the slit  the equilibrium fields $\phi^{(1)} (z)$
 and $\vartheta_i^{(1)}(z)$ contain also terms $\propto\exp(-\kappa(L-z))$ and $\propto\exp(-(L-z)/\xi)$ respectively
 (and the amplitudes are modified). The effective potential $\Psi(L)$ is obtained  by subtracting the $L$-independent 
part from $\omega_{ex}$ calculated for the equilibrium profiles.
 Neglecting $\Delta{\cal L}$  in (\ref{psi1}) we obtain  $\Psi(L)=\Psi_{DH}(L)+\Psi_C(L)$ with 
$\Psi_{DH}(L)\propto \exp(-\kappa L)$  and 
$\Psi_C(L)\propto \exp(-L/\xi)$. 

 The nonlinear terms in the EL equations (\ref{EL1}) and (\ref{EL2}) can be neglected when their magnitudes are much 
smaller than the magnitudes of the linear terms at the relevant length scales. 
The nonlinear contributions to Eqs.(\ref{EL1}) and (\ref{EL2}) can be estimated by examining $ \phi^{(1)2}(z)$ and 
$\phi^{(1)}(z)\vartheta_2^{(1)}(z)$ for $z=\xi$ and $z=\kappa^{-1}$ respectively. This is because the linear terms in
 Eqs.(\ref{EL1}) and (\ref{EL2}) decay on the length scales $\xi$ and $\kappa^{-1}$ respectively. From Eqs.(\ref{phi0}) 
and (\ref{e0}) we obtain 
$\phi^{(1)}(\xi)\propto\exp(-\xi\kappa)$ and $\vartheta_2^{(1)}(\kappa^{-1})\propto\exp(-(\xi\kappa)^{-1})$. 
Thus, the magnitudes of the correction terms depend crucially on the ratio between the correlation and the 
screening length, $\xi\kappa$. 
When $\xi\kappa\to\infty$, then $\phi^{(1)}(\xi)\to 0$ and  $\vartheta_2^{(1)}(\kappa^{-1})=O(1)$, therefore 
 we may consider linearized (\ref{EL1}), and treat  (\ref{EL2}) 
perturbatively. This case was considered in Ref.\cite{ciach:10:0} for a semi-infinite system, and in Ref.\cite{pousaneh:11:1} for a slit. On the other hand, for 
$\xi\kappa\to 0$ we have $\phi^{(1)}(\xi)=O(1)$ and $\vartheta_2^{(1)}(\kappa^{-1})\to 0$, therefore
  linearized Eq.(\ref{EL2}) can be considered. As a consequence,  $\phi^2$ in Eq.(\ref{EL1}) can be approximated by
 $\phi^{(1)2}$. In this approximation  Eq.(\ref{EL1}) takes the form of a linear inhomogeneous equation. We assume that
 this  approximation  is  reasonable as long as  $\xi\kappa<1$, and the magnitudes of $\sigma^2$ and $A_i$ are comparable.

In the experiments showing unusual dependence of the effective potential $\Psi(L)$ on $T$, and consequently on  $\xi\kappa$, the relevant lengths ratio was $\xi\kappa<1$~\cite{nellen:11:0}, therefore in this work we assume $\phi=\phi^{(1)}$. 
 Eq.(\ref{EL1}) with  $\phi$ approximated by  $\phi^{(1)}$ can be easily solved analytically. 
The excess  concentration of the solvent
 in the semi-infinite geometry takes the form 
\begin{eqnarray}
\label{cop}
 \vartheta_1(z)=\Bigg[
A_1+B_1\sigma^2 \Big(f(\xi\kappa)+f_1(\kappa,\kappa\xi)\Big)
\Bigg]\exp(-z/\xi)
\\
\nonumber
-B_{1}\sigma^2f(\xi\kappa)\exp(-2\kappa z),
\end{eqnarray}
 where $B_{1}= \frac {B^* }{2\bar \rho_c}k_B T$, $B^*$ depends on the vdW interactions and
\begin{eqnarray} 
\label{f(y)}
 f(y)=\frac{y^2}{(2y)^2-1}\to_{y\to \infty} \frac{1}{4},
\end{eqnarray}
\begin{eqnarray} 
\label{f_1(y)}
 f_1(\kappa,y)=\frac{\kappa}{y+\kappa}\Bigg \{f(y)(2y-1) -(1+2\kappa)\kappa y\Bigg[\frac{(J^{-1})_{1,2}}{B^* } 
\Big( \frac{\kappa^2}{f(y)}-1 \Big )+1\Bigg]\Bigg\}.
\end{eqnarray}
The excess solvent concentration at the distance $z$ from the hydrophobic
surface with weak and strong surface charge  is shown in Fig.2 for a few values of $\xi\kappa\le 1$ 
(note that in the figure captions the length unit $a$ is re-introduced). 
In all the cases we observe excess of organic liquid close to the surface. In some cases, however, $\vartheta_1(z)$ 
is non-monotonic and changes sign for  $z_0\sim\xi$. Excess of water appears  at the distances  $z>z_0$ 
from the surface for all 
values of  $\xi\kappa\le 1$
in the case of strong surface charges. For  weak surface charges  excess of water appears only for   $\xi\kappa<y_0(\sigma)$; 
for  $\xi\kappa>y_0(\sigma)$ a monotonic decay of  $|\vartheta_1(z)|$ occurs, as in the uncharged system. 
Thus, the presence of the 
 surface charge can change a (weakly) hydrophobic surface to an effectively hydrophilic one if we pay attention to 
the concentration of water at sufficiently large distances from the wall, $z\sim\xi$.
 Change of the adsorption preferences by increased 
surface charge was observed experimentally \cite{exp,gambassi:10:0}.  We emphasize that the change 
of the adsorption preference  for small or moderate surface charges is present  only
sufficiently far from the critical point. 

The above properties can be understood by examining Eq.(\ref{cop}) for the hydrophobic surface ($A_{1}<0$). 
For simplicity we  neglect   $f_1(\kappa,\xi)$ ($f_1(\kappa,\kappa\xi)\to 0 $ for $\xi\to\infty$ and $\kappa\to 0$ i.e. 
 for $T\to T_c$ and  $\bar\rho_c\to 0$),  
 and assume $\xi\kappa>0.5$. For $\xi\kappa>0.5$ the second term in Eq.(\ref{cop}) decays faster, 
and at the length scale $\xi$ the excess of water is found when the prefactor of the first term is positive. 
 From  Eqs.(\ref{cop}) and (\ref{f(y)}) we can conclude  that for 
$T\to T_c$ (i.e. $\xi\kappa\to\infty$) the excess of water occurs    (i.e. $\vartheta_1(\xi)>0$) when 
$A_1+\frac{\sigma^2}{4}B_{1}>0
$, which leads to the condition for the surface charge $\sigma^2>4|A_{1}|/B_1$.  
When $\sigma^2<4|A_{1}|/B_1$, excess of organic liquid occurs (i.e. $\vartheta_1(\xi)<0$)  for 
$T\to T_c$.
 Since   $\sigma^2B_{1}f(\xi\kappa)$ increases substantially when  $\xi\kappa$ 
decreases (see Eq.(\ref{f(y)})),
 the  prefactor of the first term in Eq.(\ref{cop}),  $A_1+\sigma^2B_{1}f(\xi\kappa)$, changes sign  for 
 $\xi\kappa=y_0$.  Thus, for $\sigma^2<4|A_{1}|/B_{1}$ a crossover from the excess of  the
organic liquid for $\xi\kappa>y_0$ (close to $T_c$) to the excess of water  for $\xi\kappa<y_0$ (far from $T_c$) occurs 
for sufficiently large distances from the hydrophobic surface, $z\sim\xi$.

\begin{figure}
  \includegraphics[scale=0.3]{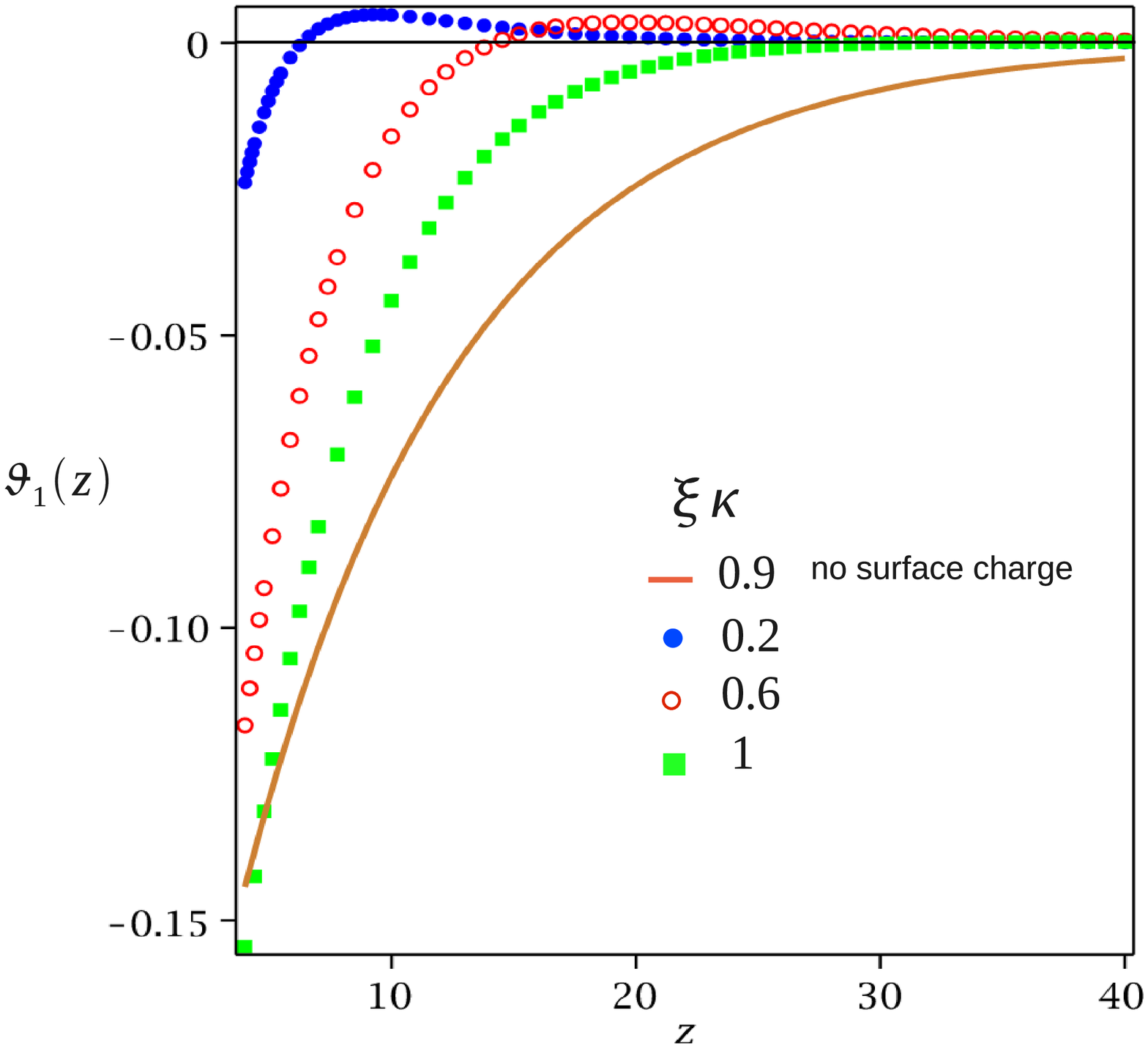}
 \includegraphics[scale=0.3]{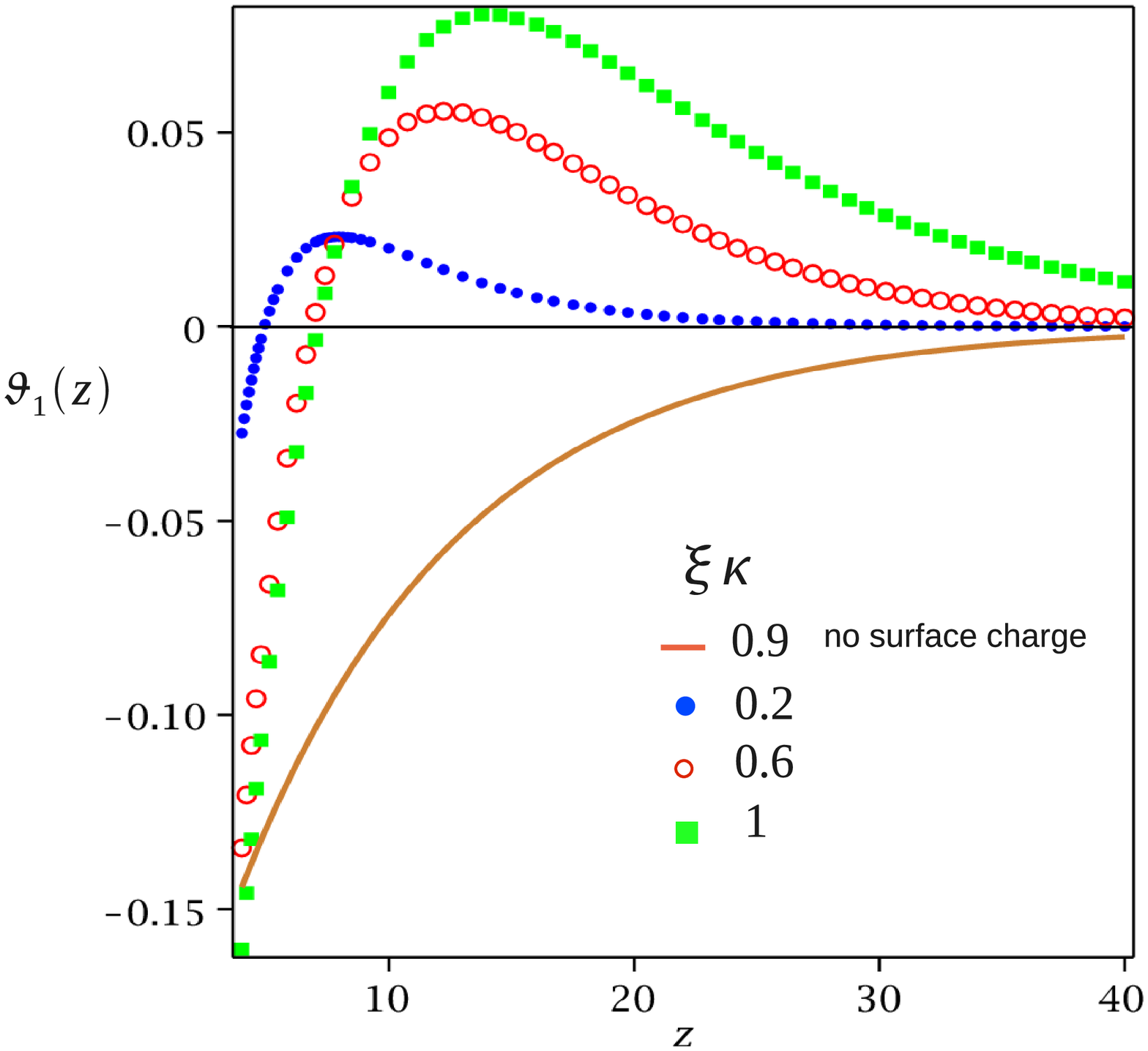}
\caption{The excess  solvent concentration $\vartheta_1(z)$ at the distance $z$  from the charged hydrophobic 
surface  for different values of $\xi\kappa$ with $\kappa=0.1a^{-1}$, where $a$ is the  
molecular diameter. The dimensionless surface charge density  (Eq.(\ref{Uel})) is (a) $\sigma=0.048/a^2 $  and  (b)
$\sigma=0.092/a^2 $,  and in Eqs.(\ref{cop}) -(\ref{f_1(y)})
 $A_1= -0.22a^{-3}$, $B^*=0.54(k_BT_ca^2)^{-1} $, $(J^{-1})_{1,2}/B^*=-12.52$
 and $\bar \rho_c=1.08\cdot 10^{-3} a^{-3}$. 
For $a=1 nm$ (approximate size of the lutidine molecule) $\bar\rho_c\approx 1.8\cdot 10^{-3} [mol/lit]$. $z$ 
and $\vartheta_1$ are in $a$ and $a^{-3}$ units respectively
and $\vartheta_1(z)>0$ for excess of water.
}
 \end{figure}

Physics behind such behavior is quite simple. The charged wall with
 no adsorption preference  attracts ions. The ions insoluble in the organic liquid attract in turn water 
molecules to this wall. The excess  number density of the hydrated ions (and thus the excess of water) appears in 
the layer of the
thickness $\sim(2\kappa)^{-1}$~\cite{barrat:03:0,israelachvili}  and depends  on the surface charge. 
The charge-neutral, hydrophobic  surface
 attracts  organic molecules. Excess of organic liquid is found in the layer of thickness $\sim \xi$, and depends 
 on the hydrophobicity of the  surface. Competition between the excess  of organic liquid and the 
excess of  water near the surface which is both hydrophobic 
 and charged  depends on $\xi\kappa$,  on the  surface charge and on
  the hydrophobicity of the wall, and  leads to the nontrivial  concentration profiles. 
\begin{figure}
  \includegraphics[scale=0.3]{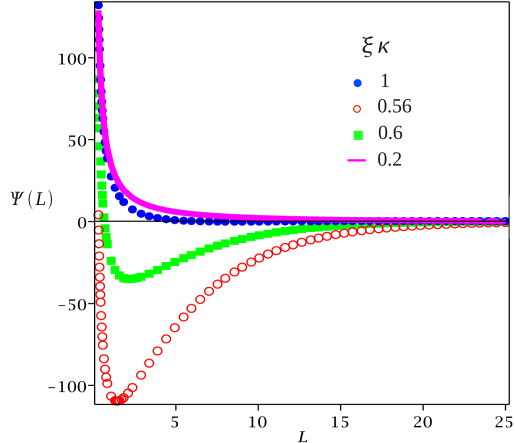}
\caption{The effective  potential per unit surface area between the charged hydrophilic and hydrophobic surfaces 
for the model system with $J^0_{1,1}=1k_BT_ca^3, J^0_{1,2}=J^0_{2,2}=0.01k_BT_ca^3$, for  $\kappa=0.1a^{-1}$
and different values of $\xi\kappa$  shown in the inset.  The surface fields (Eq.(\ref{bc0})) are
 $H_1(0)=-0.003a^{-3}, H_1(L)=0.002a^{-3}, H_2(0)=- H_2(L)=-0.5a^{-3}$, $\bar \rho_c=1.08\cdot 10^{-3} a^{-3}$, $\bar s=0$ 
 and the dimensionless surface charge density (Eq.(\ref{Uel}))
is  $\sigma_0=\sigma_L= 0.065/a^2$.  $\Psi$ is in units of $\frac{\kappa\sigma^2}{\bar\rho_c}k_BT_c$, and $L$ is in $a$ 
units, where $a$ is the  molecular diameter ($a\approx 1 nm$ for lutidine).}
\end{figure}

The Casimir potential between the walls  results from the change of the concentration near the first wall 
caused by the presence of the second wall. Let us consider vicinity of the hydrophilic wall when the weakly hydrophobic
 wall is 
present at the distance $L\sim\xi$.  The uncharged hydrophobic wall leads to depletion of water, but as discussed 
above and shown in Fig.2, in the presence of the surface charge the hydrophilic ions can 
lead to the opposite effect.  Thus, for the range of temperatures corresponding to the change of the adsorption 
preference of the weakly hydrophobic surface, the Casimir potential can be attractive. For not too large surface charge it 
could overcome the electrostatic repulsion. We calculated $\Psi(L)$  from Eqs.(\ref{psi1})-(\ref{DL})
 by inserting  the solutions of Eq.(\ref{EL1}) and linearized Eq.(\ref{EL2})
with  the boundary conditions (\ref{bc0}). The result is shown in Fig.3 for a particular model system.
 Indeed, the potential  is repulsive far from the critical point because the electrostatic repulsion dominates, becomes attractive
 and again repulsive when the critical temperature is approached. 

The above theory is derived from the microscopic statistical mechanical description by a systematic coarse-graining procedure.
 We neglected any difference in the chemical nature of the cation and the anion. Coupling between the excess concentration
 of the solvent, $\vartheta_1(z)$, and the  charge density, 
$\phi(z)$,  results first from the coupling between  $\vartheta_1(z)$ and   $\vartheta_2(z)$  in Eq.(\ref{LC}),
originating from the large difference in the solubilities of the hydrophilic ions in the two components of the solvent,
 and next from the coupling between $\vartheta_2(z)$ and $\phi(z)$  of the entropic origin (Eq.(\ref{DL})). 
 Very recently similar behavior of $\Psi(L)$ was obtained in Ref.~\cite{bier:11:0}. 
The change of the adsorption preference in Ref.~\cite{bier:11:0} results from different
 solubilities of the anion and the cation in water.
 Further studies are necessary to verify which mechanism plays the key role for the 
 experimental results reported in Ref.~\cite{nellen:11:0}.

We gratefully acknowledge discussions with A. Maciolek, U. Nellen, S. Dietrich and C. Bechinger. FP would like to thank prof. Dietrich and his group for hospitality during her stay at the MPI in Stuttgart. A part of this work  was realized within the International PhD Projects
Programme of the Foundation for Polish Science, cofinanced from
European Regional Development Fund within
Innovative Economy Operational Programme "Grants for innovation". Partial supports by the Polish Ministry of Science and 
Higher Education,  Grant No NN 202 006034, is also acknowledged.

%\bibliographystyle{prsty} 
%\bibliography{bibliography.bib}
%\end{document}

\end{document}